\documentclass[sigplan,10pt]{acmart}
\renewcommand\footnotetextcopyrightpermission[1]{}
\makeatletter
\def\@ACM@checkaffil{
    \if@ACM@instpresent\else
    \ClassWarningNoLine{\@classname}{No institution present for an affiliation}%
    \fi
    \if@ACM@citypresent\else
    \ClassWarningNoLine{\@classname}{No city present for an affiliation}%
    \fi
    \if@ACM@countrypresent\else
        \ClassWarningNoLine{\@classname}{No country present for an affiliation}%
    \fi
}
\makeatother

% ...

\usepackage{booktabs}       
\usepackage{microtype}      
\usepackage{graphicx}       
\usepackage{xspace}         
\usepackage{listings}       
\usepackage{algorithm2e}    
\usepackage{amsmath}        
\usepackage{multirow}       
\usepackage[T1]{fontenc}
\usepackage[utf8]{inputenc}
\usepackage{todonotes}      
\usepackage{tikz}           
\usepackage{mdframed}       
\usepackage{subcaption}     
\usepackage{adjustbox}
\usepackage{tabularx}
\usepackage{graphicx}
\usepackage{makecell}
\usepackage{xurl}

\newcommand{\sysname}{\textsc{Kairos}\xspace}
\newcommand{\kairos}{\textsc{Kairos}\xspace}
\newcommand{\shrikara}[1]{{{\color{purple} [Shrikara: #1]}}}

\begin{document}
\pagestyle{plain}

\title{Towards Load-Aware Prefill Deflection for Disaggregated LLM Serving}

\author{Shrikara Arun}
\affiliation{\institution{Microsoft} \country{}}
\email{t-sarun@microsoft.com}

\author{Anjaly Parayil}
\affiliation{\institution{Microsoft} \country{}}
\email{aparayil@microsoft.com}

\author{Srikant Bharadwaj}
\affiliation{\institution{Microsoft} \country{}}
\email{srikant.bharadwaj@microsoft.com}

\author{Renee St. Amant}
\affiliation{\institution{Microsoft} \country{}}
\email{reneestamant@microsoft.com}

\author{Victor Rühle}
\affiliation{\institution{Microsoft} \country{}}
\email{virueh@microsoft.com}

\begin{abstract}
Disaggregated LLM serving runs prefill and decode on separate GPU pools to keep the two phases from interfering. In practice, this creates a new asymmetry: under bursty, heavy-tailed workloads prefill nodes saturate while decode nodes have compute underutilized, and on a production-style 
A100 cluster with 2 prefill and 2 decode nodes (2P2D), we find that prefill execution accounts for only 2-23\% of P95 Time-to-First-Token (TTFT). Queuing and inter-node GPU-GPU KV-cache transfer account for the rest. 

We present a proactive prefill-deflecting scheduler that lets decode nodes serve prefill phase of requests as chunked-prefill steps interleaved with their in-flight decode batches. For each queued request, we estimate the TTFT it would see on the prefill node, and on every decode node, search for the largest chunk schedule that keeps in-flight decodes within their Time-Between-Tokens (TBT) SLO, and deflect when the decode path helps tail latency.
 Because the prefill phase of deflected requests runs in place on the decode node, the inter-node KV transfer is eliminated. Implemented on vLLM and evaluated on production-style traces with DeepSeek-V2-Lite, our approach reduces P95 TTFT by upto 81\% and raises SLO attainment by upto 79\% over state-of-the-art disaggregated schedulers, at sub-millisecond per-request routing cost.

\end{abstract}

\maketitle

\section{Introduction}
\label{sec:intro}

Large language model (LLM) inference has become a dominant datacenter workload, and
serving it efficiently at scale demands careful management of both latency and throughput. Modern production systems increasingly use disaggregated inference: prefill and decode are executed on separate GPU node pools, allowing each phase to scale independently and eliminating the GPU contention that arises when both phases compete on the same hardware~\cite{splitwise2024, distserve2024, mooncake2025}. 

\textbf{Prefill nodes become the critical bottleneck.}
In a disaggregated cluster, prefill nodes bear the full cost of processing incoming prompts before any token is generated. Under realistic workloads where prompt lengths are heavy-tailed and request arrivals are bursty, prefill nodes saturate and requests queue while decode nodes have idle compute capacity. Meanwhile, decode nodes sit underutilized. The decode phase itself is memory-bound, leaving some compute on the table and disaggregated serving never utilizes this headroom. 
Previous works such as Splitwise~\cite{splitwise2024}, DistServe~\cite{distserve2024}, Mooncake~\cite{mooncake2025}, and TetriInfer~\cite{tetriinfer2024} all share this design. On a cluster with 2 prefill and 2 decode nodes (2P2D), running representative production-style workloads, we find that prefill execution itself accounts for only 2-23\% of P95 TTFT with the rest being prefill queuing wait and inter-node GPU-GPU KV-cache transfer (\S\ref{sec:motivation}, Table~\ref{tab:mot-ttft-breakdown}).\
The result is that queue size grows on prefill nodes, especially during bursts of arrivals, while decode node compute which could be used to alleviate the pressure is underutilized. Rerouting the prefill phase of requests to the decode node also has the added benefit of removing the need for KV cache transfer, since the cache is built directly on the decode node. This is especially effective during bursts of requests, when transfer is choked. 



\textbf{Decode nodes can absorb prefill work, if routed with TBT-awareness.}
A decode node is assigned only the output token generation phase by design of disaggregation, and is not configured to function as a Sarathi-style \cite{sarathi2024} chunked-prefill colocated serving. However, the assignment of prefill requests to the decode node must be such that the utilization of idle compute is not at the cost of worsening service for existing decodes by violating their time-between-tokens (TBT) Service Level Objective (SLO) or by significantly worsening TTFT due to chunking overhead. Sarathi handles the competing TTFT and TBT SLOs by choosing a suitable chunk size, the total number of tokens in a batch and prioritizing decodes over prefills. This chunk size is fixed throughout execution in Sarathi-style serving. In reality, the state of the decode node may change drastically over time due to the bursty nature of arrivals, and thus may require different chunk sizes at different times \cite{agrawal2024medha}.
 The key question is: given an incoming request, can we deflect it to a decode node without violating SLOs?

\noindent Answering this question requires three pieces of information that are all computable at scheduling time: (1) the expected TTFT if the request stays on the prefill node, given current queue depth and request order, (2) the time to complete prefill phase of the request on decode node for a sequence of safe chunk sizes (3) the TBT headroom available on each decode node. \sysname{} computes all three and makes the deflecting decision in less than a millisecond. 

\textbf{\sysname{}: time-aware prefill deflection.}
We present \\ \sysname{}, a proactive scheduling framework for disaggregated LLM serving that reroutes prefill phase of requests to decode nodes when doing so improves tail TTFT without violating TBT SLO. For each request in the prefill phase, \sysname{} estimates its TTFT under the current prefill schedule. Concurrently, for every decode node in the cluster, it sweeps over a set of candidate chunk size sequence (chunk sizes for each step of the prefill phase) and models resulting per-step latency. For each (node, chunk size sequence) pair, \sysname{} checks two conditions: (i) TBT safety: does adding a chunked prefill to the decode node's batch increase any in-flight decode request's step latency beyond its TBT SLO? and (ii) TTFT effect: is the projected TTFT for serving a request on the decode node less than $\alpha$ times the projected TTFT on the prefill node? If both conditions hold, \sysname{} deflects prefill phase of a request to the decode node and executes it at a specific chunk-size schedule as decided by \sysname{}; otherwise the request gets routed to a prefill node. The parameter $\alpha \geq 1$ is a single operator-tunable knob: $\alpha = 1$ deflects only when the decode path is strictly faster while $\alpha > 1$ permits a bounded TTFT regression for the deflected request when doing so relieves the prefill queue for requests behind it (all within the request's own TTFT SLO). 
The design requires only one-time deployment-time profiling, no changes to the model, and a light per-request scheduling computation at the prefill dispatcher.

\textbf{Contributions} This paper makes the following contributions:
\begin{itemize}

  \item \textbf{A motivating characterization of prefill-node bottlenecks}
    (\S\ref{sec:motivation}). We show on a normalized production trace 
    that under bursty arrival patterns, prefill queuing and KV cache transfer dominates P95 TTFT even at moderate load, while decode nodes carry spare compute capacity. We also show that static policies based on thresholds  leave performance on the table. 
  \item \textbf{A TBT safety model for mixed prefill-decode batches}
    (\S\ref{sec:design}). We derive an analytical model that predicts per-step latency on a decode node as a function of in-flight decode batch size, KV cache occupancy, and added prefill chunk size. This model enables the chunk-size sweep at low cost and without live inference trials.
  \item \textbf{A time-aware deflection decision algorithm} (\S\ref{sec:design}). For each request in the prefill phase, \sysname{} (i) estimates TTFT on the prefill node, (ii) sweeps (decode node, chunk size schedule) pairs to find TBT-safe configurations, and (iii) deflects if the best decode-node TTFT is below $\alpha$ times the prefill-node TTFT. 

  \item \textbf{A full system implementation on vLLM} (\S\ref{sec:impl}). \sysname{} is implemented on top of vLLM 0.18.1 \cite{vllm2023} with approximately 2000 lines of Python code. The per-request scheduling overhead is less than 1 ms.

  \item \textbf{An evaluation}  (\S\ref{sec:eval}) on production derived trace from Company X, showing P95 TTFT improvement of upto 81\%, and upto 79\% higher SLO attainment under bursty load compared to state-of-the-art disaggregated schedulers.

\end{itemize}

\noindent
Code for \sysname{} is available at \\ \url{https://github.com/sudokara/Kairos}.

\section{Background}
\label{sec:background}

\subsection{LLM Inference: Prefill and Decode}

Modern LLM inference splits each request into two computational phases with very
different resource profiles.

\textbf{Prefill} processes the full input prompt in a single forward pass, computing
attention over all input tokens in parallel. It is \emph{compute-bound}
, and its duration scales primarily with prompt length. The output of the prefill is the KV
cache (key/value activations for every layer and token) plus the first generated token.

\textbf{Decode} generates output tokens autoregressively, one per forward pass, attending
over all previously generated tokens. It is \emph{memory-bandwidth} bound: each step reads the full KV cache but performs little new compute. Per-step 
latency typically depends on batch size and aggregate KV-cache size.

Prefill is short and compute-intensive while decode is long and memory-intensive. Running
both on the same GPU forces them to share resources that they each under use, which is the
basic motivation for disaggregated serving.

\subsection{Disaggregated Serving}

Disaggregated serving runs prefill and decode on dedicated GPU node pools, transferring the KV between instances when a request transitions from prefill to decode. 
The canonical designs Splitwise \cite{splitwise2024}, DistServe \cite{distserve2024}, Mooncake \cite{mooncake2025}, and TetriInfer \cite{tetriinfer2024}  differ in placement policy, KV-cache management, and hardware heterogeneity, but all share the assumption that a request that arrives at a prefill node stays there until prefill completes, after which its KV cache is shipped to a decode node. 
For disaggregated serving, KV cache transfer is on the critical path, and TTFT includes processing on prefill node, KV cache transfer and the return of the first token from the decode node \cite{dynamo2025kvcache}.

\sysname{} relaxes this assumption. We allow a request's prefill phase to execute on a decode node as chunked-prefill steps interleaved with the node's ongoing decode batches, exploiting decode-side compute slack that is typically unused. The remainder of this section defines the terms we use throughout the paper. We validate that decode nodes are underutilized and can accept chunked prefill steps within TBT SLO in \S\ref{sec:motivation}, and detail our design in \S\ref{sec:design}.

\subsection{Chunked Prefill}

Chunked prefill processes a request's prompt in bounded-size chunks across multiple batch steps rather than in a single forward pass. The maximum number of tokens processed per step is the \emph{chunk size} $\chi$. Sarathi \cite{sarathi2024} introduced chunked prefill to mitigate prefill-decode interference in colocated (aggregated) serving. By capping $\chi$ per step, decode steps in the same batch see bounded latency inflation, preserving inter-token latency for in-flight decodes. \sysname{} uses chunked prefill as the mechanism for executing deflected requests on decode nodes, but the per-request, state-aware $\chi$ selection it requires is what sets it apart from Sarathi's fixed-$\chi$ colocated setting (see \S\ref{sec:related}).

\subsection{Terminology}
\label{sec:terms}

We use the terms in the following table throughout the paper:
\begin{table}[h]
  \centering
    \label{tab:terms}
  \small
  \setlength{\tabcolsep}{4pt}
  \renewcommand{\arraystretch}{1.15}
  \begin{tabularx}{\linewidth}{lX}
    \toprule
    \textbf{Term} & \textbf{Definition} \\
    \midrule
    TTFT & Time-to-first-token: latency from request arrival to first output token delivered to user. \\
    TBT  & Time-between-tokens: per-step decode latency seen by in-flight requests. \\
    SLO  & Service-level objective; a per-metric latency target (e.g., 4\,s P95 TTFT, 70\,ms TBT). \\
    Chunked prefill & Processing a prompt in bounded-size chunks across multiple batch steps. \\
    Chunk size ($\chi$) & Maximum tokens processed per batch step. \\
    Deflection & Routing a request in the prefill phase to a decode node to execute via chunked prefill. \\
    $\tau$ & TBT SLO \\
    $\alpha$ & TTFT margin for the deflection condition: deflect when $\widehat{\mathit{TTFT}}_{\mathrm{dec}} \leq \alpha \cdot \widehat{\mathit{TTFT}}_{\mathrm{pf}}$, with $\alpha \geq 1$. \\
    $\beta$ & TBT error margin when calculating chunk schedule, $T_{step} \leq \beta \times \tau$ \\
    \bottomrule
  \end{tabularx}
\end{table}

\section{Empirical Study}
\label{sec:motivation}

Every experiment in this section uses request traces derived from a large-scale production LLM serving deployment. We replay all traces against DeepSeek-V2-Lite with a comparable architecture (Mixture of Experts (MoE)) to the proprietary model in the deployment. Token counts are normalized to DeepSeek-V2-Lite's 32K context window while preserving the original prompt-to-completion ratio. After normalization, prompt token counts (P25 / median / P90) are 2{,}071 / 3{,}569 / 10{,}336 and completion token counts are 21 / 50 / 163, and median prefill-to-decode (P:D) ratio is $69.7$. This places the workload in the regime where disaggregated serving is expected to pay off. NVIDIA Dynamo~\cite{dynamo2025} recommends disaggregation once the P:D ratio exceeds $8{:}1$ (assuming RDMA availability).

\textbf{Workload stratification:}
We stratify requests from the production trace into high ($>$P90) and low ($<$P25) buckets for both prompt and completion lengths, yielding four workloads (Table~\ref{tab:workloads}) with Poisson arrivals: LPLD, LPHD, HPLD, HPHD. To capture realistic temporal dynamics we add a fifth workload, Bursty, constructed directly from the long tailed production token distribution with a Gamma distributed inter-arrival process. These five workloads map to real-world scenarios such as conversational (LPHD), code completion (HPLD), retrieval augmented summarization (HPHD)  and let us isolate the effect of each phase on cluster behavior.

Unless noted, all experiments use the cluster described in \S\ref{sec:impl}: 2~prefill + 2~decode nodes (2P2D), each an NVIDIA A100 80\,GB.\footnote{We compared 2P2D against 1P3D and 3P1D on the Bursty workload: 1P3D saturates the single prefill node (median prefill queue $>$2$\times$ 2P2D), while 3P1D is capacity-limited on the single decode node, where KV-cache occupancy forces transfer back-pressure and inflates end-to-end TTFT. 2P2D balances both pressures and is the configuration where the prefill$\rightarrow$decode imbalance we study is clearest; we revisit alternative ratios in \S\ref{sec:eval}.}

\begin{table*}[t]
  \centering
  \caption{\textbf{Workload characteristics.} Each row is one of the five workloads used throughout \S\ref{sec:motivation} and is from normalized production trace of Company X. }
  \label{tab:workloads}
  \small
  \begin{tabular}{llllll}
    \toprule
    \textbf{Workload} & \textbf{Abbrev.} & \makecell{Median \\ prompt} & \makecell{Median \\ completion} & \makecell{RPS} &  \makecell{Median \\ P:D ratio}\\
    \midrule
    Low prompt, low decode       & LPLD   & 842    & 15  & 22 & 61.46 \\
    Low prompt, high decode      & LPHD   & 998    & 191 & 20 & 4.99 \\
    High prompt, low decode      & HPLD   & 14{,}013 & 9 & 1 & 1482.89 \\
    High prompt, high decode     & HPHD   & 13{,}968 & 198 & 1 & 66.53 \\
    Long-tailed, bursty          & Bursty & 3{,}569 & 50 & 4  & 69.67 \\
    \bottomrule
  \end{tabular}
\end{table*}

\subsection{P95 TTFT Breakdown}
\label{sec:mot-q1}

From \S\ref{sec:background}, we know that prefill is compute-bound and decode is memory-bandwidth-bound. We measure the extent to which this difference is exhibited across the different workloads. We also examine a breakdown of P95 TTFT to understand the main bottlenecks driving tail latency. 

We run each workload of Table~\ref{tab:workloads} to steady state on the 2P2D cluster and collect (i) per-node SM, Tensor Core, and HBM bandwidth utilization via NVIDIA Nsight Systems (Table~\ref{tab:mot-util}) and (ii) per-request P95 TTFT breakdown into prefill queue wait, prefill execution, KV-cache transfer, and decode queue wait
(Table~\ref{tab:mot-ttft-breakdown}).

\begin{table*}[t!]
  \centering
  \caption{\textbf{Steady-state utilization on prefill vs.\ decode instances.}
  SM and TC are active-cycle \%; BW is DRAM bandwidth (R\,=\,read, W\,=\,write)
  expressed as \% of the A100's 2\,TB/s peak. Mean $\pm$ s.d.}
  \label{tab:mot-util}
  \small
  \begin{tabular}{lcccccc}
    \toprule
    & \multicolumn{3}{c}{\textbf{Prefill Instance}} &
      \multicolumn{3}{c}{\textbf{Decode Instance}} \\
    \cmidrule(lr){2-4}\cmidrule(lr){5-7}
    \textbf{Workload} & SM (\%) & TC (\%) & BW R/W (\% peak) & SM (\%) & TC (\%) & BW R/W (\% peak) \\
    \midrule
    LPLD   & 76.3 $\pm$ 7.5 & 43.2 $\pm$ 5.4 & 1.58 / 0.18 & 15.5 $\pm$ 2.1  & 1.9 $\pm$ 0.2 & 0.63 / 0.05 \\
    LPHD   & 75.0 $\pm$ 7.9 & 42.7 $\pm$ 5.1 & 1.49 / 0.19 & 26.4 $\pm$ 13.8 & 5.6 $\pm$ 4.8 & 0.96 / 0.05 \\
    HPLD   & 83.9 $\pm$ 2.0 & 48.6 $\pm$ 1.8 & 0.97 / 0.28 &  7.5 $\pm$ 0.5  & 1.2 $\pm$ 0.1 & 0.28 / 0.05 \\
    HPHD   & 82.5 $\pm$ 6.7 & 47.7 $\pm$ 4.4 & 0.97 / 0.27 & 16.8 $\pm$ 2.6  & 2.1 $\pm$ 0.2 & 0.64 / 0.05 \\
    Bursty & 79.6 $\pm$ 8.4 & 45.8 $\pm$ 5.7 & 1.08 / 0.25 & 16.4 $\pm$ 3.8  & 2.0 $\pm$ 0.4 & 0.66 / 0.05 \\
    \bottomrule
  \end{tabular}
\end{table*}

\textbf{Claim 1: utilization is complementary and accentuated by some workloads.}
Across every workload, prefill SM utilization is 75-84\% while decode SM utilization never exceeds 27\%. Within that pattern, both sides show a mild but consistent workload dependence. On the prefill side, long-prompt workloads (HPLD, HPHD) push SM utilization to 83-84\% compared to the 75-76\% of short-prompt workloads because the per-step GEMMs grow with sequence length and operate closer to roofline. On the decode side, SM utilization tracks the concurrent decode batch size and cache length: LPHD with short prefills feeding long decodes at a high RPS sustains the most simultaneous decode streams and reaches 26\% SM utilization, while HPLD with long prefills feeding short decodes at a low 1 RPS leaves the decode batch nearly empty and has only 7.5\% SM utilization. The gap is widest in HPLD (84\% vs. 7.5\%). 
Bandwidth shows a similar trend, prefill nodes use under 1.6\% of HBM peak, confirming compute-boundedness, and decode nodes use under 1\%, where the limit is not raw bandwidth but the per-step $O(K)$ KV-cache reads on small decode batches that leave SMs idle.

\textbf{Claim 2: prefill compute is not the bottleneck for tail TTFT.}
Table~\ref{tab:mot-ttft-breakdown} decomposes P95 TTFT into its four constituents. Prefill execution accounts for only 2-23\% of P95 TTFT and the remaining 77-98\% is queue wait and KV-cache transfer. 

For High P workloads (1 RPS), long prompts (approximately 14K tokens) result in significant prefill queueing delays, reaching hundreds of milliseconds per engine, even with two prefill instances. KV cache transfer further contributes over 1.5 seconds at P95, with HPHD reaching up to 3.5 seconds due to memory pressure on decode nodes.   

In Low P workloads, prefill queueing is substantially lower (P95 queue wait below 60 ms), yet end-to-end TTFT remains high (P95: 6.2s for LPLD and 7.8s for LPHD). This is primarily driven by decode-side congestion at higher request rates. KV cache transfer latency can reach up to 4.2s at P95 for LPLD. 

The bursty workload exhibits both effects simultaneously: prefill queueing spikes to 840 ms, while KV cache transfer times reach 2.1s at P95 during traffic bursts.  
In all workloads, P95 TTFT is bottlenecked by queueing + KV cache transfer and not actual prefill computation. 

\begin{table*}[htbp]
  \centering
  \caption{\textbf{P95 TTFT breakdown.} Prefill execution is 2--23\% of end-to-end
  TTFT across all five workloads; prefill queue wait and inter-node KV-cache transfer
  account for the rest. Both are scheduling artifacts that deflection can eliminate.}
  \label{tab:mot-ttft-breakdown}
  \small
    \begin{tabular}{lccccc}
      \toprule
      \textbf{Workload} & Prefill queue P95 (ms) & Prefill exec P95 (ms) & KV transfer P95 (ms) & Decode queue P95 (ms) & \makecell{End-to-end \\ TTFT P95 (ms)} \\
      \midrule
      LPLD   & 59.2  & 191.4   & 4{,}205.4 & 54.5 & 6{,}278.8 \\
      LPHD   & 54.4  & 177.0   & 1{,}428.9 & 58.6 & 7{,}876.1 \\
      HPLD   & 655.3 & 1{,}340.2 & 1{,}723.4 & 37.4 & 5{,}781.5 \\
      HPHD   & 349.6 & 1{,}222.8 & 3{,}584.6 & 55.4 & 6{,}292.0 \\
      Bursty & 840.8 & 1{,}070.1 & 2{,}175.6 & 55.6 & 7{,}893.9 \\
      \bottomrule
    \end{tabular}
\end{table*}

\begin{mdframed}
\textbf{Insight 1.} Prefill compute consumes only 2-23\% of P95 TTFT with the remaining
77-98\% being prefill queue wait and inter-node KV-cache transfer. Deflecting some prefills to decode nodes can take advantage of idle SMs and an execution path that builds the KV cache in place.
\end{mdframed}

\subsection{Q2: Is There Exploitable Slack on Decode Nodes Without Violating SLO on TBT?}
\label{sec:mot-q2}

Next,
we quantify how much spare compute on decode nodes can be exploited for prefill work before decode latency degrades beyond acceptable limits.

\textbf{Methodology:}
We perform a controlled microbenchmark in which we fix a decode batch of size $B$, with each decode request having $K$ cached tokens, where $K$ corresponds to the median prompt length of the workload (Table~\ref{tab:workloads}). We then inject chunked prefill tokens, increasing the total tokens processed per batch to $\chi$, and measure the resulting step latency increase $\Delta T_{\mathrm{step}}(\chi)$ across five successive chunks. Each successive chunk adds $\chi - B$ prefill tokens while retaining $B$ decode tokens, enabling us to isolate the incremental impact of prefill injection on decode latency.

We sweep batch size $B \in \{4, 8, 16, 32\}$, cached context length $K \in \{842, 998, 14013, 13968, 3569\}$ tokens, and total batch tokens $\chi \in \{128, 256, 512, 1024, 2048\}$ to comprehensively characterize the trade-offs.

\begin{figure*}[h]
  \centering
  \includegraphics[width=\textwidth]{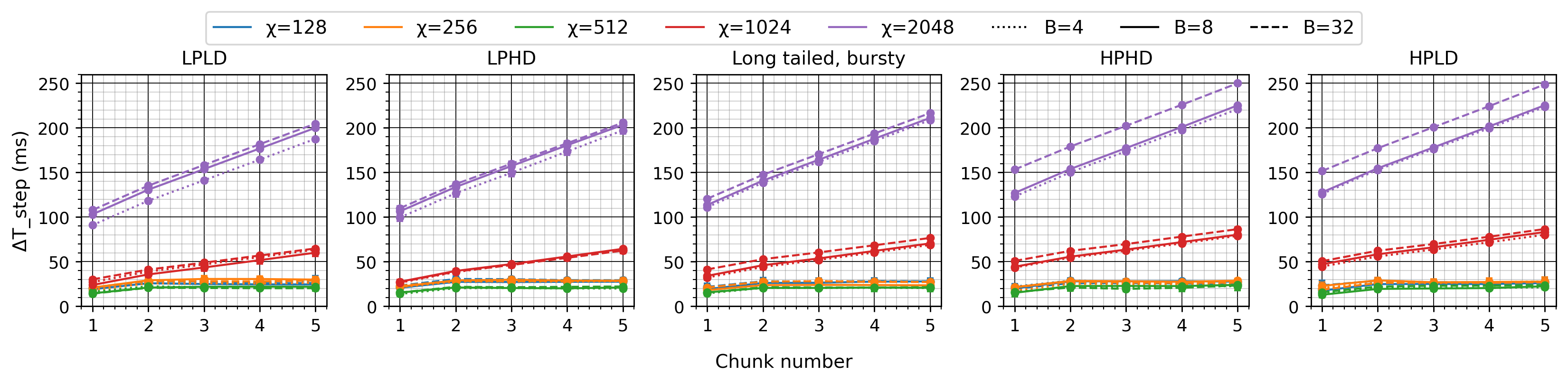}
  \caption{\textbf{Decode step latency increase from chunked prefill injection.} Step latency remains flat ($\Delta T_{\mathrm{step}} \leq$ 30\,ms) for chunk sizes up to $512$ tokens across all workloads and batch sizes.}
  \label{fig:mot-slack}
\end{figure*}

Figure~\ref{fig:mot-slack} shows the step latency increase as a function of the chunk index for different values of $\chi$ and $B$. For $\chi \leq 512$, adding prefill tokens to the decode batch increases step latency by at most 30\,ms across all tested workloads and batch sizes. This is because decode is memory bandwidth bound, the GPU stalls waiting for HBM reads, and small prefill chunks fill idle compute units without extending the memory bound critical path. However, for higher chunk sizes $\chi=1024$ and $\chi=2048$, the additional computation leads to a non negligible increase in batch time over smaller chunk sizes. We see that the absolute increase for these chunk sizes depends on the workload regime. For low prefill workloads (thus having fewer cached tokens for running decode requests), $\Delta T_{\mathrm{step}}$ for the first chunk is $\approx 30$ms. For bursty workloads and those with high prefill, this value is upto $\approx50$ms. Thus, to utilize $\chi \geq 1024$ for the first chunk, we must have knowledge of the amount of prefill cached tokens, indicating the regime we are operating in. 

As expected, higher batch sizes show higher $\Delta T_{\mathrm{step}}$. This effect is more pronounced at higher chunk sizes ($\chi\geq1024$), later chunks and workloads with high prefill cached tokens (HPHD, HPLD). 

When the prefill length exceeds $\chi - B$, multiple chunks are required. Each successive chunk adds at least $\chi - B$ cached prefill tokens, progressively increasing latency. While small chunks ($\chi \leq 512$) show minimal growth across chunks, larger chunks ($\chi = 1024, 2048$) accumulate substantial overhead, reaching up to 250\,ms by the fifth chunk for $\chi = 2048$.  

Since later chunks are more expensive, an effective strategy is to adapt chunk sizes dynamically using larger chunks early and smaller chunks later. For example, with a 50\,ms TBT headroom in bursty workloads, the first chunk can use $\chi = 1024$, while subsequent chunks must reduce to $\chi \leq 512$ to remain within budget. 

For the workloads in Table \ref{tab:workloads}, we find that actual TBT is well below the 70\,ms SLO: P50 TBT ranges from 49\,ms (HPLD) to 53\,ms (bursty), leaving 17-21\,ms of headroom at the median. Figure \ref{fig:mot-slack} shows that several chunks can be processed with this amount of headroom, using even $\chi=1024$ in some cases. Deflection allows requests to bypass the prefill queue and KV cache transfer time. Since these are significant contributors to TTFT, as shown in Table \ref{tab:mot-ttft-breakdown}, deflecting requests can dramatically reduce tail TTFT. For High P workloads (HPLD/HPHD), the advantage lies in reducing strain on prefill instances, while for Low P workloads (LPLD/LPHD) deflection offers an alternative prefill pathway. In both cases, the load on prefill instances are reduced and compute utilization of decode instances increase. Deflection also helps smooth out bursts of arrivals through this alternative prefill pathway.

\begin{mdframed}
 \textbf{Insight 2:} Small prefill chunks can be safely deflected without violating TBT but larger chunks require workload-aware control necessitating adaptive chunking. Deflection reduces queueing and KV transfer overhead, significantly improving tail TTFT.
\end{mdframed}

\subsection{Mechanism: Predicting Step Latency}
\label{sec:mot-predictors}

\S\ref{sec:mot-q2} established that decode-node slack is exploitable, but a safe chunk size depends on the in-flight decode batch, the KV occupancy, the chunk index within the prefill, and the workload regime. A practical deflection scheduler therefore cannot commit to a fixed chunk size. For every candidate (decode node, chunk size) pair it needs to predict whether the resulting step would violate TBT before dispatching. Running live trials per request is infeasible and wasteful, necessitating a closed-form predictor.

\textbf{Analytical models:}
We build three analytical step-latency predictors, one for prefill batches, one for decode batches, and one for mixed batches, plus a transfer-time model for inter-node KV migration.
All predictors require only a one-time  profiling for a model-hardware combination. On held-out traces from Table~\ref{tab:workloads}, both step-latency predictors achieve MAPE $<10\%$, which is sufficient for ranking (chunk size, decode node) pairs reliably. \sysname{} uses these predictors at scheduling time (\S\ref{sec:design}). 
More details are in \S \ref{sec:impl-estimator}.

\textbf{Simulator.}
We wrap the three predictors in an event-based simulator with support for PD disaggregation. We use this simulator to evaluate static deflection policies in \S \ref{subsec-q3-static}. 

\subsection{Q3: Do Static Deflection Policies Suffice, or Do They Cause SLO Violations Under Bursty Workloads?}
\label{subsec-q3-static}

\begin{figure*}[htpb]
  \centering
  \includegraphics[width=\textwidth]{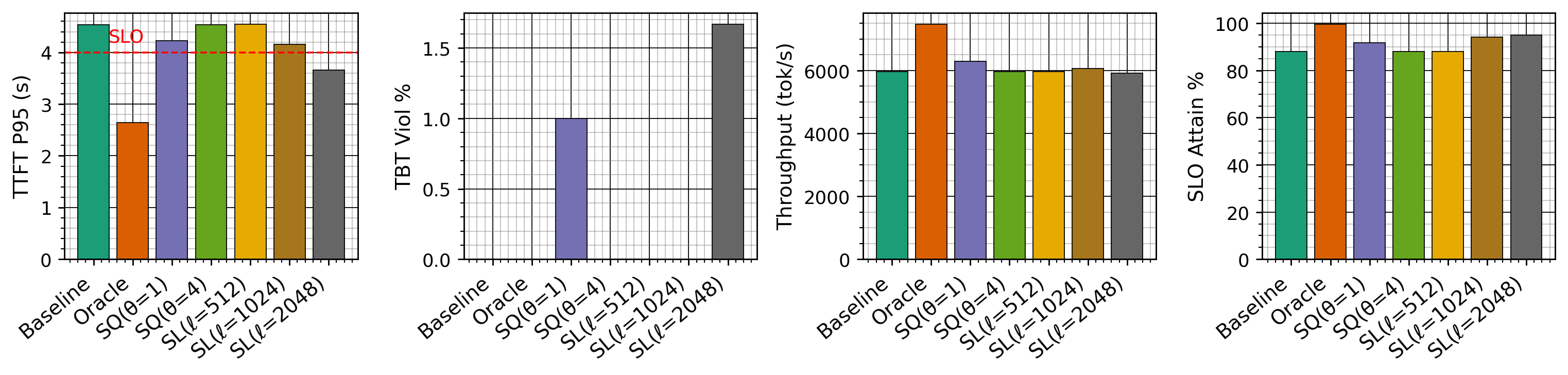}
    \caption{\textbf{Comparison of static deflection policies.} Oracle and SL($\ell=2048$) are the only policies within TTFT SLO, with Oracle having higher throughput and SLO attainment and no TBT violations. Static deflection policies perform poorly, and motivate runtime deflection decisions.}
  \label{fig:mot-static}
\end{figure*}

Given a per-request step-latency predictor (\S\ref{sec:mot-predictors}), we imagine an Oracle scheduler which sweeps every (decode node, chunk size) pair for every request in the prefill phase, using the predictors to estimate TTFT and TBT impact and deflects if a TBT-safe pair beats the prefill node. The Oracle scheduler has perfect knowledge of the system state since the simulator uses the same analytical models to determine batch time as Oracle does. Oracle is not a deployable system since it uses simulator-side state as both ground truth and decision input, but it tells us the headroom an ideal predictor-driven policy would unlock. The question we seek to answer is whether we actually need runtime per-request deflection decisions or static policies would suffice.

\textbf{Policies evaluated.}
\begin{itemize}
  \item \textbf{Baseline}: regular P/D disaggregation, no deflection.
  \item \textbf{Static-queue (SQ($\theta$))}: deflect to a decode node when prefill queue length exceeds $\theta$ on \emph{all} prefill nodes; fixed chunk size $\chi=128$.
  \item \textbf{Static-length (SL($\ell$))}: deflect every request with prompt length $<\ell$ tokens; fixed chunk size $\chi=128$.
  \item \textbf{Oracle}: per-request TBT-safe chunk-size sweep using the \S\ref{sec:mot-predictors} predictors with simulator-side state as ground truth.
\end{itemize}

\textbf{Results.}
Table~\ref{tab:mot-static-policies} summarizes every policy on the Bursty workload and describes how many requests were deflected, whether it meets the 4\,s P95 TTFT SLO, and the rate of violation of the 70\,ms TBT SLO. Oracle is the only policy that meets both SLOs simultaneously (P95 TTFT 2.6\,s, 0\% TBT violations); SL($\ell=2048$) is the only static policy within the TTFT SLO but pays for it with a 1.5\% TBT violation rate. Figure~\ref{fig:mot-static} summarizes the same.

\begin{table}[t]
  \centering
  \caption{\textbf{Static deflection policies vs.\ Oracle on the Bursty workload.} Deflection count and P95 TTFT are absolute; TBT violation rate is the fraction of decode steps exceeding the 70\,ms TBT SLO; SLO attainment is the fraction of requests meeting both the TTFT and TBT SLOs. The two failure regimes are under-deflection (SQ, SL($\ell\leq 1024$)) and over-deflection (SL($\ell=2048$)).}
  \label{tab:mot-static-policies}
  \small
  \begin{tabular}{lcccc}
    \toprule
    \textbf{Policy} & \makecell{Deflections} & \makecell{P95 TTFT\\(s) $\downarrow$} & \makecell{TBT viol.\\(\%) $\downarrow$} & \makecell{SLO attain.\\(\%) $\uparrow$} \\
    \midrule
    \makecell{Baseline\\ (no deflection)} & 0   & 4.5   & 0.0  & 88.0 \\
    SQ($\theta=1$)        & 7   & 4.2   & 1.0& 91.7 \\
    SQ($\theta=4$)        & 0   & 4.5  & 0.0  & 88.0 \\
    SL($\ell=512$)        & 3   & 4.5   & 0.0  & 88.0 \\
SL($\ell=1024$)& 151 & 4.1   & 0.0  & 94.1 \\
    SL($\ell=2048$)       & 298 & 3.6  & 1.5  & 95.1 \\
    \midrule
    \textbf{Oracle}       & 251 & \textbf{2.6} & \textbf{0.0} & \textbf{99.5} \\
    \bottomrule
  \end{tabular}
\end{table}

From Table \ref{tab:mot-static-policies}, we see that SQ($\theta=4$), SL($\ell=512$), and SL($\ell\!\leq\!1024$) all under-deflect (0-151 requests deflected), leaving TTFT gains on the table. SL($\ell=2048$) on the other hand deflects more requests (298) than Oracle by sending longer prompts to decode nodes regardless of in-flight decode state, inflating mixed-batch step latency past 70\,ms 1.5\% of the time. SQ($\theta=1$) deflects only when bursts arrive. Therefore, by the time it fires, the decode nodes already have full batches, and the prefill queue is empty, leaving the decode node to service both the burst's prefill chunks and its decodes.

Static policies involve setting inflexible thresholds, such as $\ell$ or $\theta$. This can lead to limited improvement over traditional P/D disaggregation through missed opportunities for deflection or increased violation due to limited knowledge of the state of the instances and workload characteristics. Thus, an effective policy is a per-request predictor.

\begin{mdframed}
\textbf{Insight 3.} Static policies are not ideal for deflection decisions. Under-deflection misses the bulk of Oracle's TTFT gain (up to 36\% on Bursty) and over-deflection violates TBT. The right deflection decision depends on the request, node state, and workload characteristics and necessitates per-request scheduling driven SLO-aware step-latency models.
\end{mdframed}

\section{System Design}
\label{sec:design}

Based on insights from \S \ref{sec:motivation}, we design \sysname{} with a scheduling layer that dispatches incoming requests to an appropriate node by computing a TBT safe chunked-prefill schedule on a decode node if deflection is feasible and beneficial to tail TTFT.

\subsection{Architecture Overview}
\label{sec:arch}

\begin{figure*}[t]
  \centering
  \includegraphics[width=0.6\textwidth]{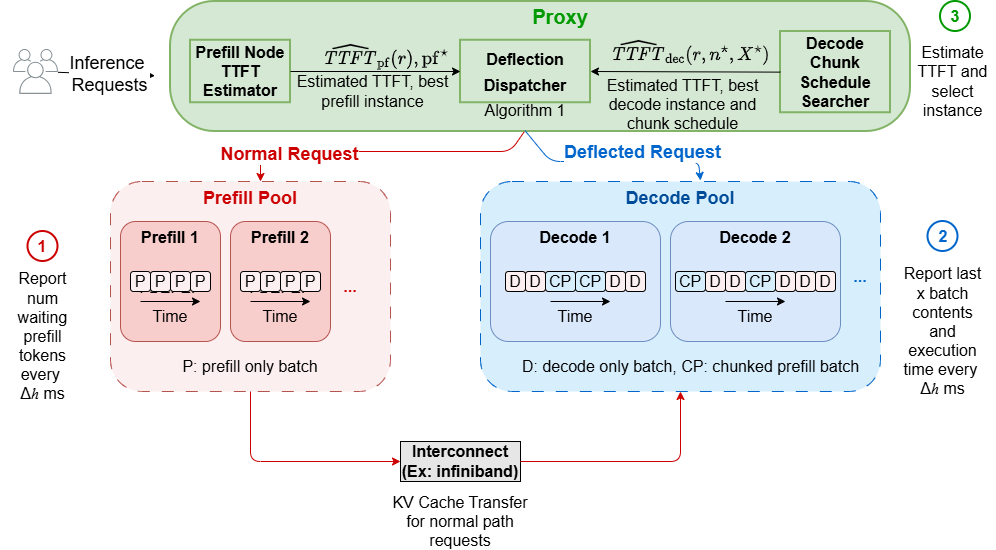}
  \caption{\sysname{} architecture.}
  \label{fig:arch}
\end{figure*}

\sysname{} adds a thin scheduling layer at the prefill dispatcher (Figure~\ref{fig:arch}). When a request joins the prefill queue, the dispatcher invokes the Deflection Decision algorithm (Algorithm~\ref{alg:deflect}) and either (a) sends the request to a prefill node (the default disaggregated path) or (b) deflects it to a decode node, where the prompt is processed as chunked-prefill steps along with the node's ongoing decode phase requests. The decision depends on two predictors - a Prefill-Node TTFT Estimator (\S\ref{sec:pf-estimator}) and a Decode-Node Feasibility Checker (\S\ref{sec:dec-checker}) which use the step-latency models described in \S\ref{sec:mot-predictors}. The model and inference kernels are unmodified and the dispatcher needs only a one-time per-hardware step-latency characterization.

\subsection{Prefill-Node TTFT Estimator}
\label{sec:pf-estimator}

For a request $r$ with prompt length $\ell_r$ arriving at the prefill dispatcher, \sysname{} estimates projected TTFT  if it stayed on the assigned prefill node:
\begin{equation}
  \widehat{\mathit{TTFT}}_{\mathrm{pf}}(r) \;=\; W_{\mathrm{queue}}(r) + T_{\mathrm{pf}}(r)
  \label{eq:ttft-pf}
\end{equation}

\textbf{Queue wait:} We assume FIFO scheduling on the prefill node, similar to the default vLLM scheduler and most production disaggregated deployments. 
Given the queue state $[\ell_1, \ldots, \ell_k]$ of prompt lengths ahead of $r$, the prefill node chunk size $\chi_{\mathrm{p}}$, the waiting time is the product of the number of chunks and the average chunk execution time considering half of outstanding prefill tokens as cached in every step:
\begin{equation}
  W_{\mathrm{queue}}(r) = \frac{\sum_{i=1}^{k} \ell_i}{\chi_{\mathrm{p}}} \times T_{step} \left( \chi_{\mathrm{p}},  \frac{\sum_{i=1}^{k} \ell_i}{2} \right)
  \label{eq:queue-wait}
\end{equation}

\textbf{Prefill execution time.} For a request of prompt length $\ell_r$:
\begin{equation}
  T_{\mathrm{pf}}(r) = \sum_{i=0}^{i=\lceil\frac{\ell_r}{\chi_p}\rceil}{T_{step}(\chi_p,  \chi_p \times i)}
  \label{eq:pf-exec}
\end{equation}
We calculate prefill execution time  as the sum of  batch execution times calculated with the prefill step latency model $T_{step}$ which is discussed next.
\subsection{Decode-Node Feasibility Checker}
\label{sec:dec-checker}

For each decode node $n$, \sysname{} checks if there is a chunked-prefill schedule
that completes request $r$'s prefill on $n$ without violating any in-flight decode's TBT and estimates the TTFT.

\subsubsection{Step-latency model}
\label{sec:step-model}

We model the per-step latency of a mixed batch on decode node $n$ as a function of the
current decode batch size $B_n$, total KV-cache occupancy $K_n$ (tokens), and the
chunked prefill size $\chi$ of that step:
\begin{equation}
  T_{\mathrm{step}}(B_n, K_n, \chi) \;=\; T_{\mathrm{step}}^{(0)}(B_n, K_n) + \Delta T_{\mathrm{pf}}(B_n, K_n, \chi).
  \label{eq:step-latency}
\end{equation}
$T_{\mathrm{step}}^{(0)}$ is the baseline decode-only step latency, $\Delta T_{\mathrm{pf}}$ is
the latency increase from injecting a $\chi-B_n$ prefill tokens into the batch. Both
terms come from the predictors in \S\ref{sec:mot-predictors}, which on held-out traces
achieve MAPE $<10\%$. Predictions are $O(1)$ via simple linear regressors. 

\subsubsection{Chunk Schedule}
\label{sec:per-chunk}

A key finding from \S\ref{sec:mot-q2} is that successive prefill chunks for the same deflected request are not equivalent. As cached prefill tokens accumulate on the decode node, $\Delta T_{\mathrm{pf}}$ grows, and the largest TBT-safe $\chi$ shrinks. \sysname{} therefore models a deflect as a schedule  \\ 
$X = [\chi^{(1)}, \chi^{(2)}, \ldots, \chi^{(m)}]$ such that $\sum_{i=1}^m \chi^{(i)} \geq \ell_r$, with each $\chi^{(i)}$ TBT-safe given the projected decode-node state at step $i$:
\begin{equation}
  T_{\mathrm{step}}\bigl(B_n, K_n + {\textstyle\sum_{j<i}}\,\chi^{(j)},\, \chi^{(i)}\bigr) \;\leq\; \beta \times \tau,
  \label{eq:tbt-safe}
\end{equation}
where $\tau$ is the TBT SLO and $\beta$ is a tunable safety factor. The schedule is constructed greedily: at step $i$ pick the largest $\chi \in \mathcal{X}$ that satisfies Eq.~\eqref{eq:tbt-safe}. The corresponding TTFT estimate on node $n$ is the sum of step latencies:
\begin{equation}
  \widehat{\mathit{TTFT}}_{\mathrm{dec}}(r, n, X) \;=\; \sum_{i=1}^{|X|} T_{\mathrm{step}}\bigl(B_n, K_n + {\textstyle\sum_{j<i}}\,\chi^{(j)},\, \chi^{(i)}\bigr).
  \label{eq:ttft-dec}
\end{equation}
If no schedule exists, node $n$ is infeasible for deflection $r$.

\subsection{Deflection Decision Algorithm}
\label{sec:algo}

Algorithm \ref{alg:deflect} composes the estimator and the feasibility checker. For request $r$, Algorithm \ref{alg:deflect} (i) estimates the prefill-node TTFT, (ii) builds a TBT-safe chunk schedule on every decode node and computes the resulting decode-node TTFT, (iii) picks the fastest feasible node, and (iv) deflects only when the $\alpha$ margin condition is met.

\RestyleAlgo{ruled}
\begin{algorithm}[t]
\small
\SetAlgoLined
\DontPrintSemicolon
\caption{\sysname{} deflection decision (per request)}
\label{alg:deflect}

\KwIn{Request $r$ with prompt length $\ell_r$,\newline
      prefill-node state: queue lengths $[\ell_1,\ldots,\ell_k]$, chunk size $\chi_{\mathrm{p}}$,\newline
      decode-node states $\{(B_n, K_n)\}_{n=1}^{N_d}$,\newline
      chunk-size candidates $\mathcal{X}=[\chi_1 > \cdots >\chi_C]$,\newline
      TBT SLO $\tau$; deflection margin $\alpha \geq 1$}
\KwOut{$(\mathrm{decision},\, n^{\star},\, X^{\star})$}
\BlankLine

\textbf{// 1. Prefill-node TTFT estimate}
$\widehat{\mathit{TTFT}}_{\mathrm{pf}}(r) \;=\; W_{\mathrm{queue}}(r) + T_{\mathrm{pf}}(r)$ \tcp*{Eq. \ref{eq:ttft-pf}}
\BlankLine

\textbf{// 2. Build a TBT-safe schedule per decode node}
\BlankLine
$\mathcal{F} \gets \emptyset$\;
\ForEach{decode node $n \in \{1,\ldots,N_d\}$}{
  $X \gets [\,]$;\quad $K \gets K_n$;\quad $\text{served} \gets 0$\;
  \While{$\text{served} < \ell_r$}{
    \tcp{first $\chi \in \mathcal{X}$ s.t.\ Eq.~\ref{eq:tbt-safe}}
    $\chi \gets \textsc{LargestSafe}(B_n, K, \mathcal{X}, \tau)$ 
    \BlankLine
    \lIf{$\chi = 0$}{\textbf{break}\tcp*[f]{node $n$ infeasible}}
    $X \gets X \,\Vert\, [\chi]$;\quad $K \gets K + \chi$;\quad $\text{served} \gets \text{served} + \chi$\;
  }
  \If{$\text{served} \geq \ell_r$}{
    $\widehat{\mathit{TTFT}}_{\mathrm{dec}}(n) \gets \sum_{\chi^{(i)} \in X} T_{\mathrm{step}}(B_n, K_n{+}\!\!\sum_{j<i}\chi^{(j)},\chi^{(i)})$ \tcp*{Eq.~\ref{eq:ttft-dec}}
    $\mathcal{F} \gets \mathcal{F} \cup \{(n, X)\}$\;
  }
}
\BlankLine

\textbf{// 3. Pick fastest feasible decode node}
\lIf{$\mathcal{F} = \emptyset$}{\KwRet $(\textsc{Keep},\,-,\,-)$}
$(n^{\star}, X^{\star}) \gets \arg\min_{(n,X) \in \mathcal{F}} \widehat{\mathit{TTFT}}_{\mathrm{dec}}(n)$\;
\BlankLine

\textbf{// 4. Deflect only if margin holds}
\BlankLine
\eIf{$\widehat{\mathit{TTFT}}_{\mathrm{dec}}(n^{\star}) \leq \alpha \cdot \widehat{\mathit{TTFT}}_{\mathrm{pf}} \text{ } \&\& \text{ } \widehat{\mathit{TTFT}}_{\mathrm{dec}}(n^{\star}) \leq TTFT_{slo}$ \tcp*{Eq. \ref{eq:deflection-condition}}}{
  \KwRet $(\textsc{Deflect},\, n^{\star},\, X^{\star})$
}{
  \KwRet $(\textsc{Keep},\,-,\,-)$
}
\end{algorithm}

\textbf{The deflection condition.} The dispatcher deflects when
\begin{equation}
\begin{split}
    \widehat{\mathit{TTFT}}_{\mathrm{dec}}(r, n^{\star}, X^{\star}) &\leq \alpha \cdot \widehat{\mathit{TTFT}}_{\mathrm{pf}}(r) \text{ and } \\ 
    \widehat{\mathit{TTFT}}_{\mathrm{dec}}(r, n^{\star}, X^{\star}) &\leq TTFT_{slo}
\end{split}
  \label{eq:deflection-condition}
\end{equation}

with $\alpha \geq 1$ an operator-tunable knob. $\alpha = 1$ deflects only when the decode path is similar or faster and values $\alpha > 1$ permit a bounded TTFT regression for $r$ in exchange for relieving the prefill queue which is useful when the request itself is already comfortably within its TTFT SLO but its presence in the prefill queue is hurting requests behind it. We choose $\alpha=1.3$.

\textbf{Complexity} The worst-case complexity of the algorithm is 
  $O\left(\lceil \frac{N_d \times \ell_r}{\chi_{min} - \text{max\_concurrency}}\rceil + N_p\right)$
 per request, where $N_d$ is the number of decode nodes, $\ell_r$ is the prompt length of the incoming request, $\chi_{\min}$ is the smallest candidate chunk size and $N_p$ is the number of prefill nodes. 
We measure that the scheduling delay is under 1ms on a single CPU core.
 
 textbf{Correctness} By construction, the algorithm never returns
$(\textsc{Deflect},\, n^{\star},\, X^{\star})$ for a schedule that the step-latency model
predicts will violate the TBT SLO. 



\section{Implementation}
\label{sec:impl}

\sysname{} is implemented on top of vLLM v0.18.1 \cite{vllm2023} which supports disaggregated prefill/decode through a KV transfer API. \sysname{} adds approximately 2000 lines of new code across the profiling for step latency models, the  Deflection Dispatcher (which contains the analytical models and the deflection algorithm), the modifications to the scheduler on the decode nodes and plugins to report node state to the Deflection Dispatcher.

\subsection{Step-Latency Profiling}

At deployment time, a one-time profiling script collects data for the analytical models by running production traces at various request rates against the target model hardware combination. The step latency and batch contents are recorded at each step and used to train the analytical models. Total profiling time is $\approx$30 minutes per model/hardware combination.

\subsection{Analytical Batch Time Estimator Models}
\label{sec:impl-estimator}

The estimators run as separate threads on the Deflection Dispatcher. Each batch execution time estimator is implemented as scikit-learn Linear Regression models. 
Node state is reported by vLLM plugins that run on each node which periodically send node state to the Deflection Dispatcher.
\begin{itemize}
  \item Prefill step latency is modeled as a function of the number of tokens in the batch and the number of cached KV tokens already on the engine. 
  \item Decode step latency is modeled as a function of the number of tokens in the batch (sum of decode tokens plus any injected prefill chunk) and the cached KV tokens.
  \item For mixed batches, step latency is modeled as the sum of pure decode step latency and additional prefill time based on total chunk size and cached tokens from the data for the microbenchmark to plot Figure \ref{fig:mot-slack}.  
  \item KV-cache transfer time is computed from bytes transferred and link bandwidth, following the Splitwise~\cite{splitwise2024} simulator.
\end{itemize}

\subsection{Deflection Dispatcher}

The Dispatcher is implemented as a Python service. It stores decode node state ($B_n$, $K_n$) and prefill node state (remaining prefill tokens) in an in-memory map, which is refreshed by reporting from nodes every $\Delta_h = 100ms$. Between heartbeats, the dispatcher marks decode nodes where it has already sent requests in the preifll state as unavailable to prevent TBT violations. The Dispatcher houses the deflection decision loop (Algorithm \ref{alg:deflect}), which completes in $<$1 ms for our setup in \S \ref{sec:eval}. 

\subsection{Chunked Prefill Scheduler on Decode Nodes}

vLLM's scheduler already supports chunked prefill as a batching strategy. When a request whose prefill phase is pending is sent to the decode node, \sysname{} sends the chunk schedule for the request in the KV parameters. 
The underlying scheduling policy remains the same, but at each step, the max tokens in the batch is governed by the corresponding chunk size in the chunk schedule of the request currently in its prefill phase.

\section{Evaluation}
\label{sec:eval}

We evaluate \sysname{} against two baselines to answer the following questions:
\textbf{Q1} Does \sysname{} reduce tail TTFT while maintaining TBT within SLO?
\textbf{Q2} Does \sysname{} improve output token throughput and overall SLO attainment?
\textbf{Q3} How many requests does \sysname{} deflect?

\subsection{Experimental Setup}
We run our experiments on Azure Machine Learning Compute Clusters \cite{aml}, using Standard\_ND96amsr\_A100\_v4 nodes.
Each node is equipped with 8$\times$ Nvidia A100-SXM4-80GB GPUs connected through 
{NVLINK 3.0, an AMD EPYC 7V12 CPU with 64-Cores
along with 900 GiB of host memory and 1.7 TiB disk storage. Internode communication is through InfiniBand \cite{infiniband} and Azure Accelerated Networking \cite{azure_accl}.} 


\subsubsection{Model, Trace and SLO:}
We evaluate \sysname{} using DeepSeek-v2-Lite \cite{deepseekv2}, a mixture of experts model with 16B total and 2.4B active parameters on the Bursty trace from Table \ref{tab:workloads}, which is the most realistic of our traces since it is not the result of any stratification. Based on the model and hardware combination, we use P95 TTFT $\leq 4$s and P95 TBT $\leq 70$ms (both must hold simultaneously) as SLO targets.

\subsubsection{Baselines:}
\begin{itemize}
  \item \textbf{PD Disaggregation}: Traditional DistServe style disaggregation \cite{distserve2024}, all prefilling done on prefill nodes.
  \item \textbf{TaiChi}: unified aggregation/disaggregation with differentiated GPU instances and flowing-decode scheduling \cite{taichi2025}.
\end{itemize}

\begin{figure}[htbp]
  \centering
  \includegraphics[width=\columnwidth]{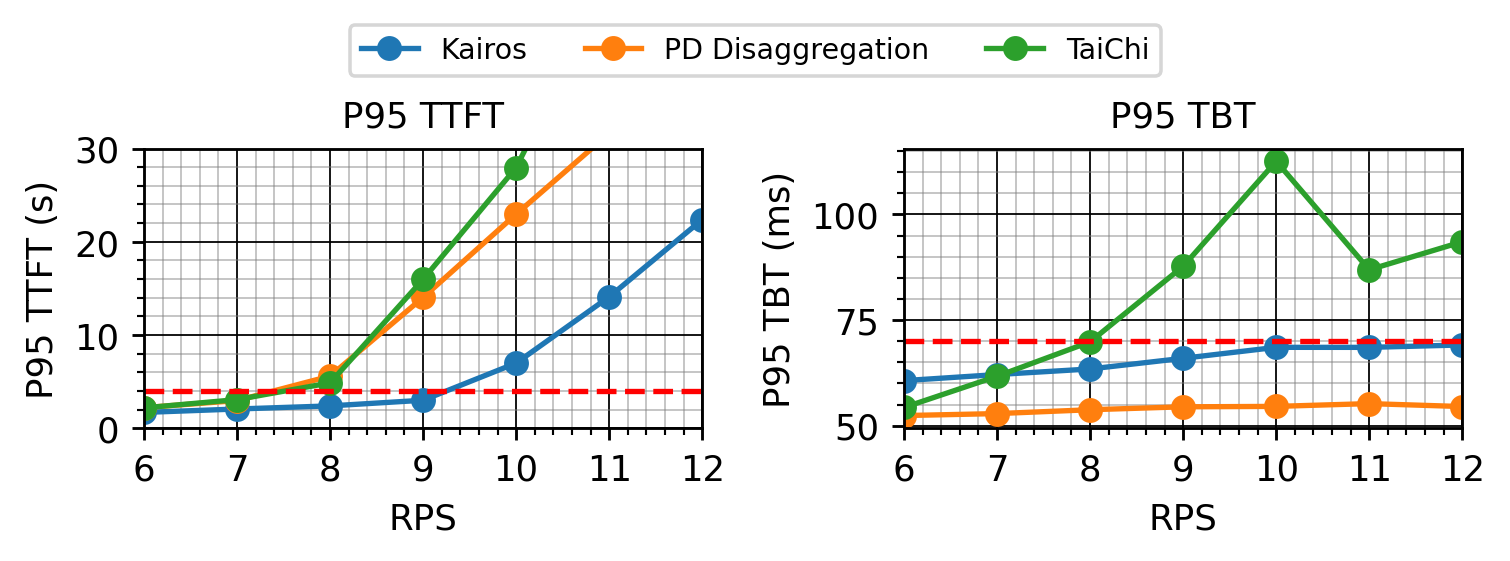}
  \caption{\textbf{L:P95 TTFT and R:P95 TBT vs. request rate.} \sysname{} stays within SLOs at a higher RPS than baselines. TaiChi violates both TTFT and TBT SLO at high request rates.}
  \label{fig:eval-e1}
\end{figure}
\subsection{Q1:  Does \sysname{} reduce tail TTFT while maintaining TBT within SLO?}
\label{sec:eval-e1}



Figure \ref{fig:eval-e1} shows P95 TTFT and TBT for each methods on the Bursty trace for RPS ranging from 6-12.
\textbf{P95 TTFT (Figure \ref{fig:eval-e1}, left)}
On the Bursty trace, \sysname{} stays within SLO upto 9 RPS, while PD disaggregation and TaiChi can only manage to do so until 7 RPS. At lower request rates, \sysname{} is similar to PD disaggregation since there is lower queuing to be skipped by deflection, while the difference rises at higher RPS. Taichi shows relatively faster increase in TTFT with RPS compared to PD disaggregation and \sysname{}. 

\textbf{P95 TBT (Figure \ref{fig:eval-e1}, right)}
\sysname{} maintains zero TBT violations by construction, and is always below the 70ms SLO. P95 TBT for \sysname{} is 5-10ms higher as it uses TBT headroom to deflect requests to the decode node. The increase in TBT is higher at higher RPS as deflections get more costly, even if they are fewer in number, due to the increase in batch size and cached KV tokens. TaiChi only manages to stay within the TBT SLO until 8 RPS and P95 TBT value crosses 100ms at 10 RPS. This shows the effectiveness of state-aware algorithm by \sysname{}. As decode nodes fill up, the safe chunk size $\chi^*$ shrinks, and \sysname{} tends to use smaller chunk size or chooses not to deflect at all.
\subsection{Q2: Does \sysname{} improve output token throughput and overall SLO attainment?}
\label{sec:eval-e2}

 Next, we measure output token throughput and SLO attainment rate to see effectiveness beyond a per request level using identical experiment setup as in \S \ref{sec:eval-e1},

\begin{figure}[tbp]
  \centering
  \includegraphics[width=\columnwidth]{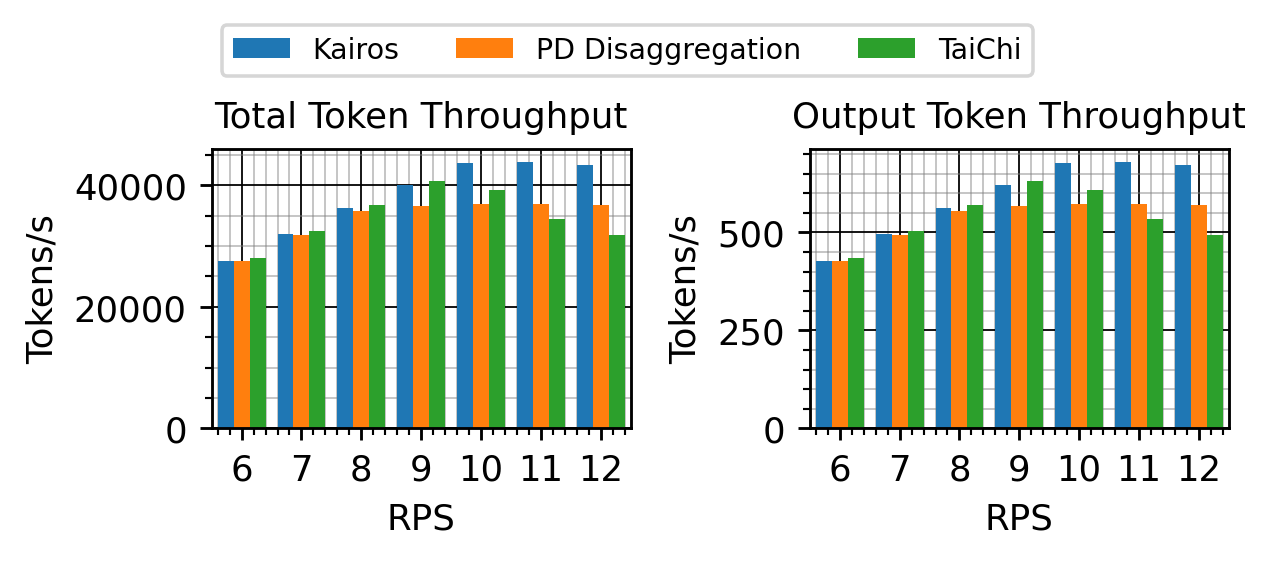}
  \caption{\sysname{} has comparable or higher throughput than baselines.}
  \label{fig:eval-throughputs}
\end{figure}

\textbf{Throughput (Figure \ref{fig:eval-throughputs})}
For both total token throughput and output token throughput, \sysname{} shows similar values to TaiChi and PD disaggregation at low RPS and higher than both baselines at higher RPS. The deflection of requests allows those subset of requests to start decoding faster, while only slowing down existing decodes marginally. Both PD disaggregation and TaiChi saturate at a lower output token throughput, with TaiChi even worsening after 9 RPS.

\textbf{SLO Attainment (Figure \ref{fig:eval-slo-deflections}, left)}
\sysname{} sustains 100\% joint SLO attainment up to 9 RPS cluster load, whereas PD Disaggregation and TaiChi attain 100\% up to 7 RPS. 
TaiChi even performs better than PD Disaggregation at RPS values of 8 and 9, however, its worsening TBT makes it fall below PD Disaggregation beyond 10 RPS. 
Degradation after the knee point is also more graceful for \sysname{}, as seen in the figure, with the drop in attainment being sharper for PD disaggregation after 7 RPS and for TaiChi after 8 RPS. 
This translates to 28.5\% higher sustainable request rate under the same SLO target.

\subsection{Q3: How many requests does \sysname{} deflect?}
\label{sec:eval-e4}

\begin{figure}[tbp]
  \centering
  \includegraphics[width=
  \columnwidth]{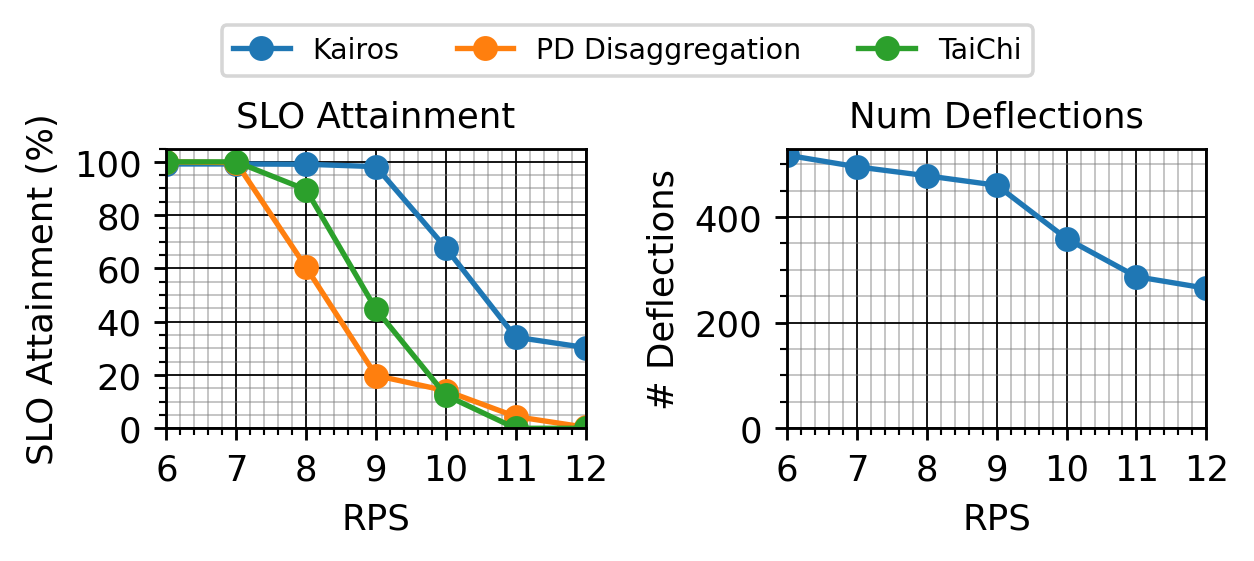}
  \caption{\textbf{Left}: \sysname{} has the best SLO attainment \% at all request rates. \textbf{Right}: Number of deflections by \sysname{} at various RPS for the Burst trace. Higher request rates see fewer deflections due to limited headroom caused by increased batch size and cached tokens.}
  \label{fig:eval-slo-deflections}
\end{figure}

From the experiment in \S \ref{sec:eval-e1}, we measure how many requests are deflected by \sysname{} at every request rate for the Burst trace. From Figure \ref{fig:eval-slo-deflections} (right), we see that at 6 RPS, \sysname{} deflects more than 500 requests to the decode node, and number of deflection reduces to 265 at 12 RPS.  

At higher request rate, more queuing at the prefill nodes  increases the potential TTFT improvement of a deflection. However, the high RPS also causes the decode nodes to be running at higher batch sizes and with more cached tokens, limiting the amount of TBT headroom that \sysname{} can exploit by deflecting a request. This limited headroom makes the safe chunk size $\chi^{\star}$ smaller, or even zero, leading to fewer deflections at higher request rates to avoid violating the TBT of running decode requests.

\section{Limitations and Future Work}
\label{sec:discussion}


\textbf{Error from analytical models:}
Error in the decode node analytical model could result in actual TBT being higher than predicted, potentially violating SLO. Error in the prefill analytical model does not put SLO at risk, and only makes a deflection decision non-optimal with slight TTFT regression for the request. The request will still meet SLO by design of Eq. \eqref{eq:deflection-condition}. With the $<10\%$ MAPE of our analytical models, we do not find any violations due to prediction errors, and better analytical models would only improve our results. 
Future works could include detecting rise in error rates to trigger retraining of the analytical models, and support for non-FIFO scheduling policies such as prefix-aware or priority.

\textbf{Stale state information at dispatch time:} Decode nodes report their state $(B_n, K_n)$ periodically (every $\Delta_h = 100ms$ in our implementation). The dispatcher uses this state to decide if a deflection is feasible, and limits each decode node to a single deflected request to prevent a single decode node from being overloaded with deflected requests. For this to become a bottleneck, the RPS must exceed $\frac{|\text{decode nodes}|}{\Delta_h} \geq 20$ which is much higher than the RPS our SLO supports. Moreover, this RPS limit can be improved by scaling the number of decode nodes or more frequent decode state reporting. Future work can include updating decode node state within dispatch until new state arrives to remove the single deflection request limit.
A decode node whose batch grows after a deflection is accepted reports its updated $(B_n, K_n)$ at the next heartbeat ($\Delta_h = 100$\,ms in our implementation). Already-dispatched requests are not migrated back. Future work could consider modeling arrival rate and characteristics of requests coming from the prefill nodes to decode nodes.


\textbf{Effect on KV cache and multi-turn:}
\sysname{}'s deflection path eliminates the KV transfer cost because the KV cache is built directly on the decode node. However, if a request has a prefix already cached on a prefill node, deflecting it to a decode node loses that cached prefix. \sysname{} currently does not account for prefix cache residency in its TTFT estimate.

\textbf{Centralized dispatcher as a potential bottleneck:}
At very high request rates, the dispatcher's per-request sweep could become a CPU bottleneck. For such deployments, the dispatcher can be sharded by node groups (a set of prefill and decode nodes are governed by a dispatcher instance), sacrificing global optimality for scalability.

\section{Related Work}
\label{sec:related}

A wide variety of works have targeted optimizing LLM inference, including optimized kernels \cite{kernel-1, kernel-2, podattention2025}, KV cache management \cite{vllm2023, sglang2024}, batching \cite{sarathi2024} and SM partitioning \cite{nexus2025, muxwise2025}. However, these works focus on single node inference optimization.  
Access to multiple nodes leads to more ways to route requests, store weights, load balance and parallelize. SplitWise \cite{splitwise2024} and DistServe \cite{distserve2024} introduced prefill/decode disaggregation by characterizing the differing compute/memory boundedness of prefill and decode phases. Mooncake \cite{mooncake2025} treats the KV cache as a first-class distributed object with a global index, enabling prefix reuse and fine-grained migration. TetriInfer \cite{tetriinfer2024} addresses disaggregated scheduling on heterogeneous GPU fleets. 

TaiChi \cite{taichi2025} proposes a unified PD aggregation-disaggregation architecture with different kinds of instances (prefill-heavy vs.decode-heavy) and configurable sliders that adapt the ratio between instances and their chunk sizes to optimize goodput under TTFT/TPOT SLO combination. 
It uses latency shifting, i.e. selectively reallocating GPU resources from SLO-satisfied requests to at-risk ones, orchestrated by flowing-decode scheduling and length-aware prefill scheduling.
On the other hand, \sysname{} identifies prefill nodes as the bottleneck during bursts, and performs a per-request TBT-safe deflection decision instead of modifying global cluster configuration. 

PPD \cite{ppd2026} also performs deflection to the decode node, but does so for Turn 2+ requests in a multi-turn setting, since these requests with high cached token counts slow down decode nodes less than Turn 1 prompts. While \sysname{} and PPD share the same idea of deflection to the decode node, \sysname{} is more general since any request can be deflected if there is headroom, and also incorporates run-time monitoring of nodes to make deflection decisions that do not violate TBT. 
\begin{table}[t]
  \centering
  \caption{\kairos{} vs.\ other related systems}
  \label{tab:related-ea}
  \small
  \begin{tabular}{p{1.3cm}p{1.6cm}p{1.8cm}p{1.7cm}}
    \toprule
    \textbf{System} & \textbf{Mechanism} & \textbf{Scope} & \textbf{Target} \\
    \midrule
    Sarathi \cite{sarathi2024}        & Token chunked prefill   & Intra-node         & Decode stall \\
    Nexus \cite{nexus2025}           & Intra-GPU SM partition  & Intra-GPU          & Phase interference \\
    PPD \cite{ppd2026}             & Append-prefill routing  & Cross-node (multi-turn) & TTFT Turn~2+ \\
    TaiChi \cite{taichi2025}         & Aggreg.+disagg.\ sliders & Cluster config    & Goodput under any SLO \\
    Splitwise \cite{splitwise2024}       & Static PD split         & Cluster topology   & Phase separation \\
    DistServe \cite{distserve2024}      & Placement optimization  & Cluster provision  & Node ratio \\
    \midrule
    \textbf{\kairos{}} & \textbf{Per-req TBT-safe chunk sweep} &
      \textbf{Cross-node deflection} & \textbf{Prefill bottleneck} \\
    \bottomrule
  \end{tabular}
\end{table}

\section{Conclusion}
\label{sec:conclusion}


This paper presents \sysname{}, a proactive deflection framework that utilizes unused decode node compute to serve deflected requests. For each incoming request, \sysname{} estimates its projected TTFT through the regular disaggregated and deflection paths, and deflects request to a decode node with available capacity when doing so meets SLO of incoming and existing requests.



Our evaluation on production traces from Company X on DeepSeek-v2-Lite shows that \sysname{} reduces P95 TTFT by upto 81\%, and can improve SLO attainment rate by upto 71\% for realistic bursty workloads while adding negligible per-request scheduling overhead and respecting the same SLO.


\newpage
\bibliographystyle{ACM-Reference-Format}
\bibliography{references}

\end{document}